\documentclass[twocolumn,tighten]{aastex631}

\usepackage{nameref}
\usepackage{hyperref}
\usepackage{booktabs}
\usepackage{longtable}
\usepackage{amsmath}
\usepackage{amssymb}
\usepackage{float}
\usepackage{array}
\usepackage{threeparttable}
\usepackage{color}
\usepackage{graphicx}
\usepackage{multirow}
\usepackage{tabularx}
\usepackage[figuresright]{rotating}
\usepackage[version=4]{mhchem}

\graphicspath{{./}{figures/}}

\begin{document}

\title{Modeling Atmospheric Ion Escape from Kepler-1649 b and c over Time}

\correspondingauthor{Haitao Li, Lianghai Xie}
\email{lihaitao@nssc.ac.cn, h.li6@exeter.ac.uk, xielianghai@nssc.ac.cn}

\author[0000-0001-6395-1936]{Haitao Li}
%\altaffiliation{Co-first authors} 
\affiliation{National Space Science Center, Chinese Academy of Sciences, Beijing 100190, China}
\affiliation{Department of Physics and Astronomy, Faculty of Environment Science and Economy, University of Exeter, EX4 4QL, UK}
\affiliation{University of Chinese Academy of Sciences, Beijing 100049, China}
%\email{lihaitao@nssc.ac.cn}

\author[0000-0002-8990-094X]{Chuanfei Dong}
\affiliation{Department of Astronomy, Boston University, Boston, MA 02215, USA}
%\email{dcfy@bu.edu}

\author[0000-0001-9635-4644]{Lianghai Xie}
\affiliation{State Key Laboratory of Solar Activity and Space Weather, National Space Science Center, Chinese Academy of Sciences, Beijing 100190, China}
%\email{xielianghai@nssc.ac.cn}

\author[0009-0002-8332-9225]{Xinyi He}
\affiliation{National Space Science Center, Chinese Academy of Sciences, Beijing 100190, China}
\affiliation{University of Chinese Academy of Sciences, Beijing 100049, China}

\author[0009-0006-7877-1835]{Laura Chin}
\affiliation{Department of Astronomy, Boston University, Boston, MA 02215, USA}

\author[0000-0000-6025-6663]{Xinke Wang}
%\altaffiliation{Co-first authors}
\affiliation{National Space Science Center, Chinese Academy of Sciences, Beijing 100190, China}
\affiliation{University of Chinese Academy of Sciences, Beijing 100049, China}

\author[0000-0002-8609-3599]{Hong-Liang Yan}
\affiliation{CAS Key Laboratory of Optical Astronomy, National Astronomical Observatories, Beijing 100101, China}
\affiliation{Institute for Frontiers in Astronomy and Astrophysics, Beijing Normal University, Beijing 102206, China}
\affiliation{School of Astronomy and Space Science, University of Chinese Academy of Sciences, Beijing 100049, China}

\author[0009-0000-2934-4067]{Jinxiao Qin}
\affiliation{Department of Physics, Hebei Normal University, Shijiazhuang 050024, China}
\affiliation{Guo Shoujing Institute for Astronomy, Hebei Normal University, Shijiazhuang 050024, China}

\author[0000-0001-6707-4563]{Nathan Mayne}
\affiliation{Department of Physics and Astronomy, Faculty of Environment Science and Economy, University of Exeter, EX4 4QL, UK}

\author[0000-0002-1624-3360]{Mei Ting Mak}
\affiliation{Department of Physics and Astronomy, Faculty of Environment Science and Economy, University of Exeter, EX4 4QL, UK}
\affiliation{Atmospheric, Oceanic, and Planetary Physics, Department of Physics, University of Oxford, Oxford OX1 3PU, UK}

\author[0000-0002-7071-5437]{Nikolaos Georgakarakos}
\affiliation{Division of Science, New York University Abu Dhabi, PO Box 129188, Abu Dhabi, UAE}
\affiliation{Center for Astrophysics and Space Science (CASS), New York University, Abu Dhabi, PO Box 129188, Abu Dhabi, UAE}

\author[0000-0002-4997-0847]{Duncan Christie}
\affiliation{Max-Planck-Institut f\"{u}r Astronomie, K\"{o}nigstuhl 17, 69117 Heidelberg, Germany}

\author{Yajun Zhu}
\affiliation{State Key Laboratory of Solar Activity and Space Weather, National Space Science Center, Chinese Academy of Sciences, Beijing 100190, China}
\affiliation{University of Chinese Academy of Sciences, Beijing 100049, China}

\author[0000-0003-4609-4519]{Zhaojin Rong}
\affiliation{Key Laboratory of Earth and Planetary Physics, Institute of Geology and Geophysics, Chinese Academy of Sciences, Beijing 100029, China}
\affiliation{College of Earth and Planetary Sciences, University of Chinese Academy of Sciences, Beijing 100049, China}
\affiliation{Mohe Observatory of Geophysics, Institute of Geology and Geophysics, Chinese Academy of Sciences, Mohe Heilongjiang 165303, China}

\author[0009-0000-9313-5251]{Jinlian Ma}
\affiliation{National Space Science Center, Chinese Academy of Sciences, Beijing 100190, China}
\affiliation{University of Chinese Academy of Sciences, Beijing 100049, China}

\author[0000-0003-4585-589X]{Xiaobo Li}
\affiliation{Key Laboratory of Particle Astrophysics, Institute of High Energy Physics, Chinese Academy of Sciences, 19B Yuquan Road,
Beijing 100049, China}

\author{Shi Chen}
\affiliation{National Space Science Center, Chinese Academy of Sciences, Beijing 100190, China}
\affiliation{University of Chinese Academy of Sciences, Beijing 100049, China}

\author{Hai Zhou}
\affiliation{National Space Science Center, Chinese Academy of Sciences, Beijing 100190, China}
\affiliation{University of Chinese Academy of Sciences, Beijing 100049, China}



\begin{abstract}

Rocky planets orbiting M-dwarf stars are prime targets for atmospheric characterization, yet their long-term evolution under intense stellar winds and high-energy radiation remains poorly constrained. The Kepler-1649 system, hosting two terrestrial exoplanets orbiting an M5V star, provides a valuable laboratory for studying atmospheric evolution in the extreme environments typical of M-dwarf systems. In this Letter we show that both planets could have retained atmospheres over gigayear timescales. Using a multi-species magnetohydrodynamic model, we simulate atmospheric ion escape driven by stellar winds and extreme ultraviolet radiation from 0.8 to 4.0~Gyr. The results reveal a clear decline in total ion escape rates with stellar age, as captured by a nonparametric LOWESS regression, with \ce{O+} comprising 98.3\%–99.9\% of the total loss. Escape rates at 4.0 Gyr are two to three orders of magnitude lower than during early epochs. At 0.8 Gyr, planet b exhibits 3.79× higher \ce{O+} escape rates than planet c, whereas by 4.0 Gyr its \ce{O+} escape rate becomes 39.5× lower. This reversal arises from a transition to sub-magnetosonic star–planet interactions, where the fast magnetosonic Mach number, $M_f$, falls below unity. Despite substantial early atmospheric erosion, both planets may have retained significant atmospheres, suggesting potential long-term habitability. These findings offer predictive insight into atmospheric retention in the Kepler-1649 system and inform future JWST observations of similar M-dwarf terrestrial exoplanets aimed at refining habitability assessments.

\end{abstract}



\section{Introduction} \label{sec:introduction}

The search for habitable exoplanets has evolved significantly in recent decades, driven by advances in observational technology and numerical modeling \citep{2013Sci...340..577S, kaltenegger2017characterize, 2022ExA....54.1275B}. Initially focused on broad surveys of planetary systems, the focus has shifted toward characterizing terrestrial exoplanets within the habitable zone \citep[HZ, the region around a star where liquid water might persist on the surface of rocky planets][]{gonzalez2005habitable, 2023ApJ...948L..26H}. The importance of this endeavor lies in its potential to identify environments capable of supporting life, a question that has become increasingly urgent with the discovery of thousands of exoplanets \citep{2018AsBio..18..663S, 2018AsBio..18..709C, 2021ARA&A..59..291Z}. Current research frontiers emphasize atmospheric retention and evolution, particularly for rocky planets orbiting M-dwarfs, where stellar activity poses unique challenges to habitability \citep{shields2016habitability, ABC20, 2023MNRAS.525.5168M, 2023ApJ...951L..39K}.

The retention of the atmospheres of terrestrial exoplanets orbiting M-dwarfs remains a pivotal element in assessing their potential habitability \citep{2016MNRAS.459.4088O,2022ARA&A..60..159W, 2024AJ....167..196K}. Although the HZ defines regions where liquid water could theoretically exist \citep{gonzalez2005habitable}, the sustainability of an atmosphere, particularly under prolonged stellar activity, ultimately determines whether surface conditions remain biologically viable \citep{cockell2016habitability, 2020IAUS..354..461E, 2023MNRAS.518.2472R}. For M-dwarf systems, where intense stellar winds and extreme ultraviolet (EUV) fluxes persist over gigayear timescales \citep{stevenson2003planetary,2019LNP...955.....L, khodachenko2021impact, 2024ApJ...960...62E}, atmospheric erosion and evolution require time-dependent analysis to evaluate long-term planetary habitability \citep{lammer2008atmospheric}. 

Significant progress in understanding atmospheric escape has been made through studies of exoplanetary systems \citep{2019AREPS..47...67O, 2025MNRAS.537.1305B}. Atmospheric escape processes exhibit complex dependencies on stellar and planetary parameters that remain inadequately characterized for Earth-sized exoplanets \citep{2020JGRA..12527639G, 2023ApJ...951..136L}. Studies of solar system analogs such as Mars reveal orders-of-magnitude variations in atmospheric ion loss over evolutionary timescales \citep{Dong_2018, 2018Icar..315..146J}, while three-dimensional magnetohydrodynamic (MHD) simulations of exoplanets highlight the critical roles of varying stellar wind \citep{Garraffo2016,dong2017proxima}, planetary magnetic fields \citep{Garcia-Sage2017,2024A&A...688A.138P}, planetary body size \citep{Chin2024}, planetary atmospheric composition \citep{dong2020atmospheric, Lee2021ApJ}, and orbital architecture \citep{Bourrier2018,dong2019role}. Recent JWST observations further underscore the importance of characterizing the atmospheres of terrestrial exoplanets and quantifying their evolutionary pathways \citep{Santos_2023,Rackham2023,TJCI2024}, particularly in systems where stellar activity drives non-thermal atmospheric escape.

The Kepler-1649 system provides a unique laboratory for investigating these processes, hosting two terrestrial planets at different orbital distances. Kepler-1649 b (Venus-like, 0.051\,AU) and c (Earth-like, 0.088 AU) \citep{vanderburg2020habitable, coughlin2020discovery, 2021AJ....161...31K}. Despite their similar radii ($\sim 1.0$--$1.1 \, R_\oplus$), the two planets experience distinct space weather environments due to their different orbital distances from the M5V host star, likely leading to divergent atmospheric evolution pathways \citep{dong2018atmospheric}. While previous studies have characterized atmospheric loss in M-dwarf systems based on either stellar winds and radiation at the current epoch or limited evolutionary stages \citep{dong2017proxima, dong2018atmospheric, dong2020atmospheric,2020ApJ...897..101C,2023MNRAS.525.5168M}, the effects of long-term stellar evolution on non-thermal atmospheric erosion over gigayear timescales remain unexplored. This gap is particularly significant for systems like Kepler 1649, where two Earth-sized planets at distinct orbits offer a unique opportunity to study divergent atmospheric evolution pathways under sustained stellar influences.

This study aims to quantify the time-dependent ion escape from Kepler-1649 b and c to to assess their potential for atmospheric retention throughout the host star's evolution. This investigation contributes to our understanding of exoplanet habitability by characterizing atmospheric loss around M-dwarfs, a class of stars that host many known terrestrial planets. Its significance lies in bridging theoretical models with future observations, such as those enabled by the James Webb Space Telescope (JWST) \citep{Santos_2023}, which may detect atmospheric signatures like \ce{CO2} \citep{2023ApJ...956L..13M}. By assessing whether these planets can sustain atmospheres over gigayear timescales, this work addresses a key criterion for habitability and enhances the interpretation of upcoming spectroscopic observations \citep{2022AJ....163..140L, 2024ApJ...966..189Y}. While prior 1D models \citep{2018A&A...617A.107J, 2019A&A...624L..10J, 2020ApJ...890...79J, 2021E&PSL.57617197J, 2022ApJ...937...72N} emphasize cooling in hydrodynamic escape for Earth-like atmospheres, and \cite{ 2024A&A...683A.153V} predict stripping via Jeans in TRAPPIST-1, our MHD approach reveals ion-driven trends enabling retention in Kepler-1649, highlighting non-thermal mechanisms' role.

In this work, we employ a multi-species MHD model to simulate ion escape from Kepler-1649 b and c over 0.8 to 4.0 Gyr. Our approach incorporates: (1) time-dependent stellar wind parameters derived from rotational braking models \citep{2017A&A...603A..58R}, and (2) EUV flux evolution calibrated to M-dwarf activity cycles \citep{2023MNRAS.525.4008V}. This enables quantitative predictions of atmospheric erosion under realistic stellar evolutionary scenarios. 
The structure of this Letter is arranged as follows. In Section \ref{sec:setup}, we provide a detailed description of the simulation setup, including the calculation of various parameters essential for the simulations. Subsequently, we present the results obtained from the simulations in Section \ref{sec:result}. Finally, we summarize our findings based on the analysis in Section \ref{sec:discussion}.

\section{The Simulation Setup} \label{sec:setup}
In this section, we present the numerical model used for our simulations, as well as the methods used to calculate the required input parameters.

\subsection{Simulation Parameters} \label{sec:pm}
To simulate stellar winds is challenging. The latest models \citep[e.g.,][]{Cohen_2023} incorporate three-dimensional self-consistent approaches. However, these models often require knowledge of the stellar magnetic field maps as input to simulate stellar winds. In the absence of observed magnetic maps for Kepler-1649, we adopt the Parker stellar wind model to calculate the stellar wind parameters. Such assumptions have also been made in previous research \citep{2014A&A...570A..99S,2023MNRAS.525.5168M}. The calculation is performed in the spherical coordinate system, and we convert the results to Cartesian coordinates and use them for subsequent simulation (Section \ref{sec:Parker_wind} in the Appendix). The calculated data are listed in Table \ref{table1}. We calculate the stellar wind parameters and interplanetary magnetic field (IMF) of Kepler-1649 using the stellar radius (0.252~$R_{\odot}$) and mass (0.219~$M_{\odot}$) reported by \cite{angelo2017kepler}. The rotational period at different ages was calculated using empirical relations from \cite{2023MNRAS.525.5168M}. Due to the lack of direct observations of the surface magnetic field for Kepler-1649, we derive an age-dependent estimate of the surface magnetic field through interpolation based on simulation results from \cite{2023MNRAS.519.5304L}, which is essential for estimating the IMF.

We estimate the age of Kepler-1649 from MESA isochrones v1.2 \citep{2011ApJS..192....3P, 2016ApJS..222....8D}. The stellar parameters of Kepler-1649 were taken from \cite{vanderburg2020habitable}. The age of Kepler-1649 was estimated by matching its stellar parameters (with uncertainties) to a set of isochrones with a stellar mass of 0.2 $M_{\odot}$ but varying metallicities covering the range of [Fe/H]=-0.15$\pm$0.11. The best fit isochrone results in $\log_{10}\tau \sim$9.3, which corresponds to the age $\tau$ of Kepler-1649 $\sim 2.0$ Gyr.

Because the spectral type of Kepler-1649 is M5V \citep{angelo2017kepler}, we use the spectral data of GJ 551 with a spectral type of M5.5V from MUSCLES \citep{2016ApJ...820...89F} as the spectrum of Kepler-1649. We scale the spectrum at the surface of the Kepler-1649 to obtain fluxes at the positions of Kepler-1649 b and c, which are input parameters for photoionization (see Figure \ref{Fig:Kepler1649bc_spec} in the Appendix). For the different ages of Kepler-1649, we employ the relationship given by \cite{2017A&A...603A..58R} for the temporal evolution of GJ 551 X-ray and extreme-ultraviolet (XUV) flux ($F_{\rm XUV}$ in ${\rm Wm^{-2}}$, $\tau$ in Myr)

\begin{equation}
\label{eq_XUV}
F_{\rm XUV} = \left\{
\begin{array}{lcl}   
84.1\tau^{-0.71} & & {10 <\tau<300},\\
1.47& & {300<\tau<1640},\\
9.74 \times 10^4\tau^{-1.5}  & & {1640 <\tau<4800}.
\end{array} \right.
\end{equation}

Using this relationship, we obtain the evolved stellar spectrum and calculate the photoionization rates (Section \ref{sec:chemical_reactions} in the Appendix). Incorporating GJ 551's time dependent spectral data and the ionization cross sections of \ce{CO2} and $\ce{O}$, we derive the photoionization rates for different stellar ages. Combining the inverse square relationship between XUV flux and the orbital radius, we derive the photoionization rates of Kepler-1649 b and c, which are listed in Table \ref{table1}. The XUV flux is treated as wavelength-dependent flux (Equation C8), with absorption optical depth approximated via zenith-angle cosine factors and average absorption cross-sections of \ce{CO2} and $\ce{O}$ \citep{ma2013global}. This yields self-consistent ion production but may underestimate optical depth-dependent effects. Simulations begin at 0.8 Gyr post-primordial/water loss (do Amaral et al. 2022), and end at the twice the age of Kepler-1649. \ce{H+} from stellar wind enables charge exchange, modulating \ce{O+} production without contributing to planetary loss.

To define the temporal scope of our time-dependent ion escape simulations, we establish a starting point of 0.8~Gyr and an upper bound of 4.0~Gyr, reflecting both astrophysical constraints and habitability considerations. The choice of 4.0~Gyr as the endpoint is motivated by the significant uncertainty in Kepler-1649's age, yet potentially ranging up to several gigayears due to ambiguities in M-dwarf evolutionary tracks and metallicity variations \citep{2011ApJS..192....3P,2023MNRAS.525.5168M}. We adopt a timeline comparable to that of the Solar System, where the emergence of complex life and the stabilization of a modern atmosphere demanded nearly the entirety of Earth's evolutionary history. This cutoff emphasizes the critical role of prolonged atmospheric retention in habitability assessments for terrestrial exoplanets, as the development of conditions supportive of complex life likely requires billions of years of atmospheric stability \citep{2018AsBio..18..663S}. The starting time of 0.8~Gyr is set due to the limitations of the stellar wind evolution model (Section \ref{sec:Parker_wind} in the Appendix), which becomes increasingly unreliable for ages $\lesssim$0.7~Gyr \citep{2005ApJ...628L.143W,Popinchalk_2021,2023MNRAS.525.5168M}. By initiating at 0.8~Gyr, we ensure the simulation captures a physically realistic baseline for atmospheric erosion while avoiding extrapolation into poorly constrained early stellar phases.

Due to the harsh space environment to which Kepler-1649 b and c are exposed, they might undergo processes similar to Venus during the early stages of atmospheric evolution, leading to significant loss of water and $\rm H_2$, resulting in $\ce{CO2}$ and $\ce{O}$ becoming the predominant neutral constituents of their atmospheres \citep{angelo2017kepler}. Recent research on Venus Zone terrestrial planets shows Kepler-1649 b resides in the Venus Zone, with Kepler-1649 c positioned at the zone's outer boundary \citep{2014ApJ...794L...5K, 2023AJ....165..168O}. Our assumed Venus-like composition (\ce{CO2} and \ce{O} as primary neutrals, including \ce{H+}, \ce{O+}, \ce{O2+}, \ce{CO2+} ions) is appropriate for Kepler-1649 b and c given their harsh stellar environments, which likely promote water loss and \ce{CO2} dominance \citep{angelo2017kepler}. Recent studies on Kepler-1649 c indicate that the primordial atmosphere escapes 30 Myr earlier, accompanied by significant water loss \citep{2022ApJ...928...12D}. However, alternative compositions (e.g., \ce{N2}-rich or \ce{H2O}-dominated) could alter photochemistry and escape rates. Lighter species enhance escape, potentially doubling loss rates, while denser \ce{CO2} atmospheres resist erosion better due to higher scale heights and recombination rates. For example, the simulations show that the unmagnetized TOI-700 d with a 1 bar Earth-like atmosphere could be stripped away rather quickly ($<$1 Gyr), while the unmagnetized TOI-700 d with a 1 bar \ce{CO2}-dominated atmosphere could persist for many billions of years \citep{dong2020atmospheric}. Furthermore, we calculate the neutral atmosphere by scaling the neutral atmosphere of Venus following \cite{dong2017proxima}. The Venusian neutral atmosphere\footnote{The scale height of Venus $H_{\rm Venus}$ is 5.5 km for \ce{CO2} and 17 km for \ce{O}.} from \cite{ma2013global} is
\begin{align}
    [\ce{CO2}] & = 1.0 \cdot 10^{15} \cdot e^{-(z-z_0)/5.5}  \: \rm cm^{-3}, \\
    [\ce{O}]    & = 2.0 \cdot 10^{11} \cdot e^{-(z-z_0)/17}  \: \rm cm^{-3},
\end{align}
where $z_0$ = 100 km. Studies have shown that ion escape rates exhibit a weak dependence on surface pressure \citep{dong2017proxima}. Since the surface pressure is currently unknown, we assume that the surface atmospheric pressure of Kepler-1649 b and c is 1\,bar \citep{dong2017proxima}. Thus, we obtain a density that is 0.011 times the Venus value at the model lower boundary. Next, because the gravity and temperature changes, the scale heights of Kepler-1649 b and c are
\begin{align}
    H_{\rm Kepler-1649\, b} = \frac{kT_{\rm p,b}}{m g_{\rm b}},\\
    H_{\rm Kepler-1649\, c} = \frac{kT_{\rm p,c}}{m g_{\rm c}}.
\end{align}
The subscripts b and c in the above formula represent Kepler-1649 b and c respectively. The planetary parameters of Kepler-1649 b and c are taken from \cite{vanderburg2020habitable}.

%\begin{deluxetable*}{cchlDlccc}
\begin{deluxetable*}{ccccccccc}

%    \tablenum{1}
    \label{table1}
    \tablecaption{Stellar wind parameters, photoionization rates, average temperature of upper neutral atmosphere, and upstream fast magnetosonic Mach number for Kepler-1649 b and c.}
    \tablewidth{0pt}
    \tablehead{
    \colhead{Age (Gyr)} & \colhead{$N_{\rm sw}$ (cm$^{-3}$)}  & \colhead{$T_{\rm sw}$ (K)} & 
    \colhead{$V_{\rm sw}$ (km/s)} & \colhead{IMF (nT)} & \colhead{$q_{\ce{CO2}}$ (s$^{-1}$)} & \colhead{$q_{\ce{O}}$ (s$^{-1}$)}& \colhead{$T_{\rm p}$ (K)}  & \colhead{$M_{\rm f}$ }
    }
    \startdata
        \multicolumn{9}{c}{Kepler-1649 b} \\
     0.8  & 2579 & $1.84\times10^{6}$ & (-586,0,0) & (-400.92,-0.38,0) & $1.55\times10^{-4}$ & $5.81\times10^{-5}$ & 1313.25  & 2.4995  \\
     1.0  & 1891 & $1.59\times10^{6}$ & (-529,0,0) & (-365.16,-0.32,0) & $1.55\times10^{-4}$ & $5.81\times10^{-5}$ & 1313.25  & 2.2471  \\
     1.2  & 1466 & $1.41\times10^{6}$ & (-487,0,0) & (-335.82,-0.27,0) & $1.55\times10^{-4}$ & $5.81\times10^{-5}$ & 1313.25  & 2.0580  \\
     1.4  & 1182 & $1.27\times10^{6}$ & (-454,0,0) & (-312.55,-0.23,0) & $1.55\times10^{-4}$ & $5.81\times10^{-5}$ & 1313.25  & 1.9051  \\
     1.6  & 981 & $1.17\times10^{6}$ & (-427,0,0) & (-296.50,-0.21,0) & $1.55\times10^{-4}$ & $5.81\times10^{-5}$ & 1313.25  & 1.7619   \\
     1.8  & 833 & $1.08\times10^{6}$ & (-404,0,0) & (-282.62,-0.19,0) & $1.35\times10^{-4}$ & $5.04\times10^{-5}$ & 1183.50  & 1.6428  \\
     2.0  & 719 & $1.01\times10^{6}$ & (-384,0,0) & (-270.59,-0.17,0) & $1.15\times10^{-4}$ & $4.31\times10^{-5}$ & 1059.18  & 1.5379  \\
     2.2  & 630 & $9.45\times10^{5}$ & (-367,0,0) & (-260.07,-0.16,0) & $9.97\times10^{-5}$ & $3.73\times10^{-5}$ & 962.45  & 1.4498  \\
     2.4  & 558 & $8.92\times10^{5}$ & (-352,0,0) & (-250.76,-0.15,0) & $8.75\times10^{-5}$ & $3.28\times10^{-5}$ & 885.44  & 1.3714   \\
     2.6  & 499 & $8.47\times10^{5}$ & (-339,0,0) & (-242.47,-0.14,0) & $7.76\times10^{-5}$ & $2.90\times10^{-5}$ & 829.77  & 1.3028  \\
     2.8  & 450 & $8.06\times10^{5}$ & (-327,0,0) & (-235.42,-0.13,0) & $6.94\times10^{-5}$ & $2.60\times10^{-5}$ & 797.30  & 1.2387  \\
     3.0  & 409 & $7.71\times10^{5}$ & (-316,0,0) & (-229.11,-0.12,0) & $6.26\times10^{-5}$ & $2.34\times10^{-5}$ & 770.15  & 1.1804  \\
     3.2  & 374 & $7.39\times10^{5}$ & (-307,0,0) & (-223.41,-0.11,0) & $5.68\times10^{-5}$ & $2.13\times10^{-5}$ & 745.83  & 1.1312  \\
     3.4  & 344 & $7.10\times10^{5}$ & (-298,0,0) & (-218.22,-0.11,0) & $5.19\times10^{-5}$ & $1.94\times10^{-5}$ & 716.46  & 1.0836  \\
     3.6  & 318 & $6.84\times10^{5}$ & (-290,0,0) & (-213.44,-0.10,0) & $4.76\times10^{-5}$ & $1.78\times10^{-5}$ & 691.11  & 1.0413  \\
     3.8  & 295 & $6.60\times10^{5}$ & (-282,0,0) & (-208.95,-0.10,0) & $4.39\times10^{-5}$ & $1.64\times10^{-5}$ & 669.06  & 1.0002  \\
     4.0  & 275 & $6.39\times10^{5}$ & (-275,0,0) & (-204.82,-0.09,0) & $4.07\times10^{-5}$ & $1.52\times10^{-5}$ & 649.72  & 0.9641  \\
        \midrule
        \multicolumn{9}{c}{Kepler-1649 c} \\
     0.8  & 799 & $1.84\times10^{6}$ & (-640,0,0) & (-136.12,-0.20,0) & $5.27\times10^{-5}$ & $1.97\times10^{-5}$ & 721.45  & 3.3570  \\
     1.0  & 583 & $1.59\times10^{6}$ & (-581,0,0) & (-123.98,-0.17,0) & $5.27\times10^{-5}$ & $1.97\times10^{-5}$ & 721.45  & 3.1317  \\
     1.2  & 450 & $1.41\times10^{6}$ & (-537,0,0) & (-114.02,-0.14,0) & $5.27\times10^{-5}$ & $1.97\times10^{-5}$ & 721.45  & 2.9497  \\
     1.4  & 361 & $1.27\times10^{6}$ & (-501,0,0) & (-106.12,-0.12,0) & $5.27\times10^{-5}$ & $1.97\times10^{-5}$ & 721.45  & 2.7871  \\
     1.6  & 299 & $1.17\times10^{6}$ & (-473,0,0) & (-100.67,-0.11,0) & $5.27\times10^{-5}$ & $1.97\times10^{-5}$ & 721.45  & 2.6350  \\
     1.8  & 253 & $1.08\times10^{6}$ & (-448,0,0) & (-95.96,-0.10,0) & $4.57\times10^{-5}$ & $1.71\times10^{-5}$ & 679.95  & 2.4976  \\
     2.0  & 218 & $1.01\times10^{6}$ & (-428,0,0) & (-91.87,-0.09,0) & $3.91\times10^{-5}$ & $1.46\times10^{-5}$ & 640.19  & 2.3809  \\
     2.2  & 190 & $9.45\times10^{5}$ & (-410,0,0) & (-88.30,-0.08,0) & $3.39\times10^{-5}$ & $1.27\times10^{-5}$ & 608.86  & 2.2734   \\
     2.4  & 168 & $8.92\times10^{5}$ & (-394,0,0) & (-85.14,-0.08,0) & $2.97\times10^{-5}$ & $1.11\times10^{-5}$ & 581.70  & 2.1755  \\
     2.6  & 150 & $8.47\times10^{5}$ & (-380,0,0) & (-82.32,-0.07,0) & $2.63\times10^{-5}$ & $9.86\times10^{-6}$ & 559.66  & 2.0872  \\
     2.8  & 135 & $8.06\times10^{5}$ & (-367,0,0) & (-79.93,-0.07,0) & $2.36\times10^{-5}$ & $8.83\times10^{-6}$ & 541.48  & 2.0020  \\
     3.0  & 122 & $7.71\times10^{5}$ & (-356,0,0) & (-77.79,-0.06,0) & $2.13\times10^{-5}$ & $7.96\times10^{-6}$ & 526.28  & 1.9249  \\
     3.2  & 111 & $7.39\times10^{5}$ & (-345,0,0) & (-75.85,-0.06,0) & $1.93\times10^{-5}$ & $7.22\times10^{-6}$ & 513.42  & 1.8487  \\
     3.4  & 102 & $7.10\times10^{5}$ & (-336,0,0) & (-74.09,-0.06,0) & $1.76\times10^{-5}$ & $6.60\times10^{-6}$ & 502.42  & 1.7865  \\
     3.6  & 94 & $6.84\times10^{5}$ & (-327,0,0) & (-72.47,-0.05,0) & $1.62\times10^{-5}$ & $6.05\times10^{-6}$ & 490.81  & 1.7236   \\
     3.8  & 87 & $6.60\times10^{5}$ & (-319,0,0) & (-70.95,-0.05,0) & $1.49\times10^{-5}$ & $5.58\times10^{-6}$ & 480.09  & 1.6673  \\
     4.0  & 81 & $6.39\times10^{5}$ & (-312,0,0) & (-69.54,-0.05,0) & $1.38\times10^{-5}$ & $5.17\times10^{-6}$ & 470.69  & 1.6178  \\
        \midrule
        \enddata
\end{deluxetable*}

\subsection{MHD model and setup} \label{sec:MHD}
In the previous subsection, we introduce the method used to calculate the stellar wind parameters, the IMF, the XUV flux and resulting photoionization rate. In this subsection, we will describe the MHD model used for the subsequent simulations and the parameters we have adopted.

We use the 3D Block Adaptive Tree Solar-Wind Roe Up-Wind Scheme (BATS-R-US) multi-species MHD (MS-MHD) model \citep{Toth12} to simulate the stellar wind interaction with Kepler-1649 b and c. BATS-R-US has many modules for simulating different physical phenonmena. This model has been successfully
applied to simulate atmospheric ion escale for Venus-like exoplanets \citep{dong2017proxima, dong2018atmospheric, dong2020atmospheric}. The MS-MHD model solves four continuity equations for each ion species which are $\rm H^+,O^+,O^+_2,$ and $\rm CO^+_2$, and treats each ion species as a fluid, requiring one momentum equation and one energy equation \citep{ma2013global}. This model self-consistently includes photoionization, charge exchange, and electron recombination. The photochemical and photoionization treatments employed here are preliminary, prioritizing computational efficiency for long-term evolutionary simulations over detailed kinetic modeling. Future work will incorporate more comprehensive schemes. Heating is primarily chemical, with XUV contributions via photoelectron excess energy; IR heating is neglected due to the static neutral model. Due to the lack of direct observations, we assume that the atmospheric compositions of Kepler-1649 b and c are close to that of Venus \citep{angelo2017kepler}. Non-thermal electrons are incorporated via excess energy from photoionization, enhancing ionospheric heating. Cooling via \ce{O}/\ce{CO2} emission, conduction, and eddy diffusion is approximated through adopted profiles from \cite{2006P&SS...54.1425K}. We use their average temperature $T_{\rm p}$ of lower thermosphere to approximate neutral temperature changes and scale height changes accordingly in our simulations,  but uncertainties persist and future work will couple dynamic neutral models. The reaction rates used for the simulations are listed in Table \ref{table3} (Section \ref{sec:chemical_reactions} in the Appendix).

We adopt a nonuniform grid to ensure that the radial resolution inside the ionosphere is 5\,km, while the outer boundary resolution is thousands of kilometers. The angular resolution is $3^{\circ}$. For the coordinate system, the positive direction of the $x$ axis is directed from the planet to the star. The $z$ axis is perpendicular to the orbital plane, and the $y$ axis constitutes a right-hand system. The computational domain is defined by $-20 \, R_{\rm P} \leq { X} \leq 12 \, R_{\rm P}$, $-16 \, R_{\rm P} \leq { Y,Z} \leq 16 \, R_{\rm P}$, where $R_{\rm P}$ is the radius of the planet.

\section{Results} \label{sec:result}

This section presents the calculated ion escape rates for the Kepler-1649 system. We first establish the stellar age dependence of ion escape rates through nonparametric regression of data points with LOWESS method \citep{Cleveland01121979}, then examine spatial ion distribution changes in response to evolving stellar wind and radiation conditions. Comparative analysis between Kepler-1649 b and c reveals how orbital distance modulate erosion efficiency across gigayear timescales.

\begin{figure*}
	\centering 
    \resizebox{\hsize}{!}
	{\includegraphics[width=\textwidth]{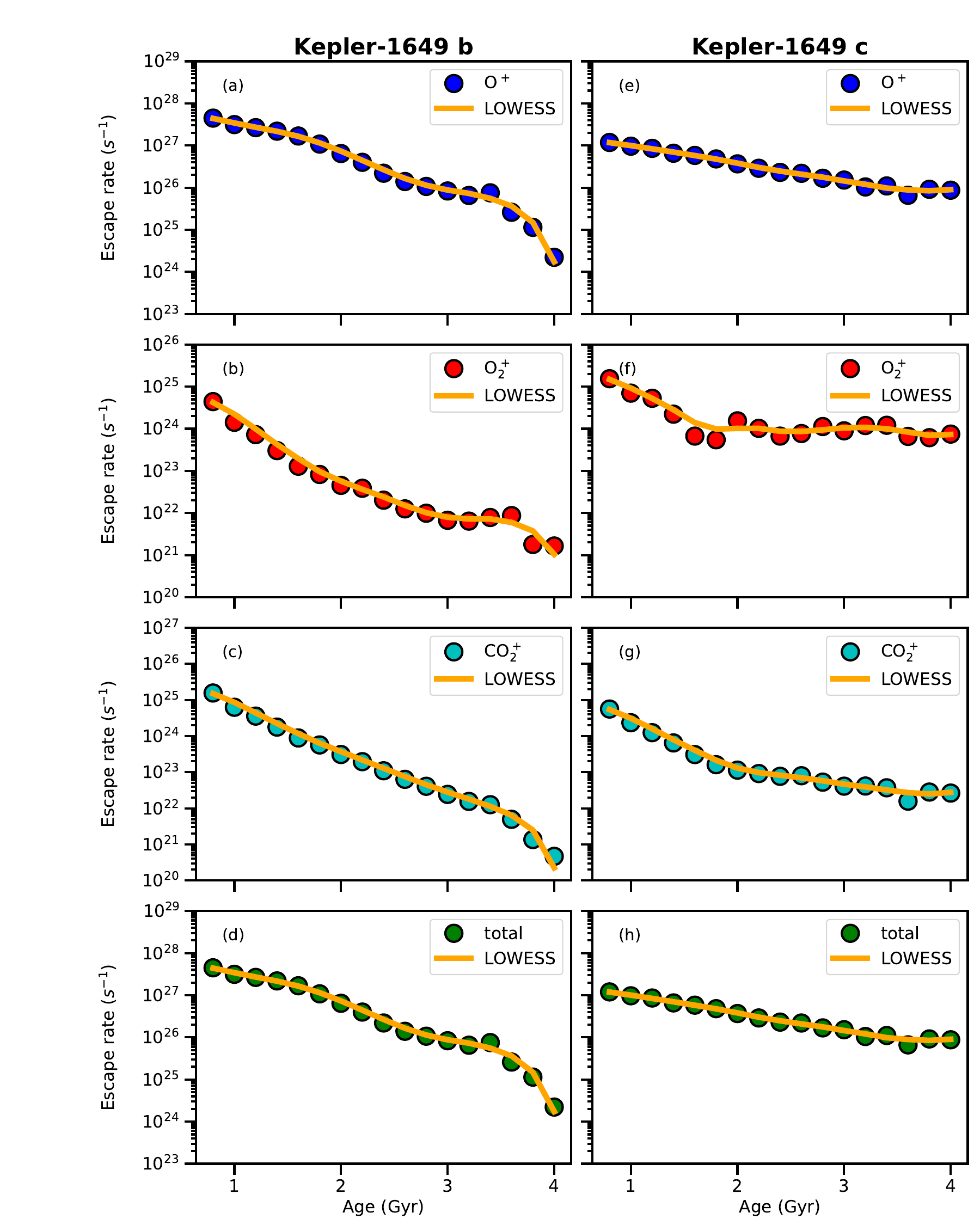}}
	\caption{Evolution of ion escape rates for Kepler-1649 b (left panels) and c (right panels) across stellar ages (0.8--4.0 Gyrs). 
Top to bottom: \ce{O+}, \ce{O2+}, \ce{CO2+}, and total ion escape rates (s$^{-1}$). The data points were nonparametrically regressed using the LOWESS method (solid lines). 
	\label{fig1}}
\end{figure*}

Figure~\ref{fig1} and Table~\ref{table4} present the temporal evolution of atmospheric ion escape rates for Kepler-1649 b and c across 0.8–4.0~Gyr, simulated under varying stellar wind and XUV radiation conditions. The ion escape rates exhibit a clear decline with stellar age, as captured by the nonparametric LOWESS regression \citep{Cleveland01121979}. This nonparametric approach provides a robust description of the trends derived from our model, avoiding assumptions about functional forms. The data reveal the total escape rates spanning four orders of magnitude ($10^{24}$–$10^{27}~\rm s^{-1}$), quantified through multi-species MHD simulations using a 10~$R_{\rm P}$ integration sphere. Both planets exhibit systematic declines in \ce{O+}, \ce{O2+}, and \ce{CO2+} escape fluxes with stellar age, with total rates decreasing from $4.49 \times 10^{27}~\rm s^{-1}$ (Kepler-1649 b) and $1.20 \times 10^{27}~\rm s^{-1}$ (Kepler-1649 c) at 0.8~Gyr to $2.22 \times 10^{24}~\rm s^{-1}$ and $8.77 \times 10^{25}~\rm s^{-1}$ for Kepler-1649 b and Kepler-1649 c at 4.0~Gyr. Table~\ref{table4} presents the calculated different ion escape rates at 17 evolutionary stages, allowing a direct comparison of absolute values and relative ion contributions.

The data in Figure~\ref{fig1} reveal a consistent decline in ion escape rates for both Kepler-1649 b and c across all species (\ce{O+}, \ce{O2+}, \ce{CO2+}, and total) over stellar ages from 0.8 to 4.0 Gyr. Escape rates for Kepler-1649 b decrease by approximately three orders of magnitude, with \ce{O+} showing the highest initial rate at $4.47 \times 10^{27}$ s$^{-1}$ and the lowest at $2.22 \times 10^{24}$ s$^{-1}$, while Kepler-1649 c exhibits also a two-order-of-magnitude reduction, with \ce{O+} rates ranging from $1.18 \times 10^{27}$ to $8.69 \times 10^{25}$ s$^{-1}$. The \ce{O+} species consistently dominates the total escape rate, contributing the majority of atmospheric loss for both planets. Total escape rates at 0.8 Gyr are $4.49 \times 10^{27}$ s$^{-1}$ for Kepler-1649 b and $1.20 \times 10^{27}$ s$^{-1}$ for Kepler-1649 c, decreasing to $2.22 \times 10^{24}$ s$^{-1}$ and $8.77 \times 10^{25}$ s$^{-1}$ at 4.0 Gyr, respectively. This pattern suggests a strong age-dependent reduction in atmospheric erosion, with Kepler-1649 b experiencing a steeper decline compared to Kepler-1649 c, reflecting a potential influence of orbital distance on long-term atmospheric retention.

Table~\ref{table4} quantifies the dominance of \ce{O+} in atmospheric loss. For Kepler-1649 b, \ce{O+} contributes $98.3\%$--$99.9\%$ of the total escape flux across all ages, while \ce{O2+} and \ce{CO2+} contribute $0.07\%$--$0.1\%$ and $0.02\%$--$0.35\%$, respectively. Similarly, \ce{O+} accounts for $98.3\%$--$99.1\%$ of Kepler-1649 c's total escape, with \ce{O2+} and \ce{CO2+} contributing $0.85\%$--$1.3\%$ and $0.03\%$--$0.46\%$. Notably, Kepler-1649 b maintains higher \ce{O+} escape rates than Kepler-1649 c at early epochs, and lower \ce{O+} escape rates than Kepler-1649 c at late epochs.

\begin{table}[h]
%	\tablenum{3}
	\centering
    \caption{The calculated atmospheric ion escape rates (of different ion species) in units of $\rm sec^{-1}$ as a function of stellar age for Kepler-1649 b and c.}\label{table4}
	\begin{tabular}{cccccc}
		\hline
		\hline
		Age (Gyr) & \ce{O+} & \ce{O2+} & \ce{CO2+} & Total \\
		\hline
		\multicolumn{5}{c}{Kepler-1649 b} \\ 
     0.8  & $4.47\times10^{27}$ & $4.41\times10^{24}$ & $1.56\times10^{25}$ & $4.49\times10^{27}$  \\
     1.0  & $3.13\times10^{27}$ & $1.42\times10^{24}$ & $6.28\times10^{24}$ & $3.14\times10^{27}$  \\
     1.2  & $2.65\times10^{27}$ & $7.26\times10^{23}$ & $3.58\times10^{24}$ & $2.65\times10^{27}$  \\
     1.4  & $2.19\times10^{27}$ & $3.03\times10^{23}$ & $1.77\times10^{24}$ & $2.19\times10^{27}$  \\
     1.6  & $1.68\times10^{27}$ & $1.29\times10^{23}$ & $8.86\times10^{23}$ & $1.68\times10^{27}$ \\
     1.8  & $1.08\times10^{27}$ & $8.21\times10^{22}$ & $5.68\times10^{23}$ & $1.08\times10^{27}$  \\
     2.0  & $6.45\times10^{26}$ & $4.53\times10^{22}$ & $3.09\times10^{23}$ & $6.45\times10^{26}$  \\
     2.2  & $3.99\times10^{26}$ & $3.86\times10^{22}$ & $1.96\times10^{23}$ & $3.99\times10^{26}$  \\
     2.4  & $2.20\times10^{26}$ & $2.01\times10^{22}$ & $1.09\times10^{23}$ & $2.20\times10^{26}$ \\
     2.6  & $1.40\times10^{26}$ & $1.25\times10^{22}$ & $6.34\times10^{22}$ & $1.40\times10^{26}$  \\
     2.8  & $1.07\times10^{26}$ & $9.87\times10^{21}$ & $4.07\times10^{22}$ & $1.07\times10^{26}$  \\
     3.0  & $8.31\times10^{25}$ & $6.71\times10^{21}$ & $2.42\times10^{22}$ & $8.32\times10^{25}$  \\
     3.2  & $6.51\times10^{25}$ & $6.42\times10^{21}$ & $1.55\times10^{22}$ & $6.52\times10^{25}$  \\
     3.4  & $7.53\times10^{25}$ & $7.82\times10^{21}$ & $1.26\times10^{22}$ & $7.53\times10^{25}$  \\
     3.6  & $2.59\times10^{25}$ & $8.68\times10^{21}$ & $4.94\times10^{21}$ & $2.60\times10^{25}$  \\
     3.8  & $1.14\times10^{25}$ & $1.79\times10^{21}$ & $1.36\times10^{21}$ & $1.14\times10^{25}$  \\
     4.0  & $2.22\times10^{24}$ & $1.65\times10^{21}$ & $4.66\times10^{20}$ & $2.22\times10^{24}$  \\
\hline
\multicolumn{5}{c}{Kepler-1649 c} \\ 
     0.8  & $1.18\times10^{27}$ & $1.54\times10^{25}$ & $5.61\times10^{24}$ & $1.20\times10^{27}$  \\
     1.0  & $9.62\times10^{26}$ & $7.00\times10^{24}$ & $2.34\times10^{24}$ & $9.72\times10^{26}$  \\
     1.2  & $8.54\times10^{26}$ & $5.29\times10^{24}$ & $1.24\times10^{24}$ & $8.61\times10^{26}$  \\
     1.4  & $6.55\times10^{26}$ & $2.22\times10^{24}$ & $6.46\times10^{23}$ & $6.58\times10^{26}$  \\
     1.6  & $5.81\times10^{26}$ & $6.70\times10^{23}$ & $3.07\times10^{23}$ & $5.82\times10^{26}$  \\
     1.8  & $4.79\times10^{26}$ & $5.47\times10^{23}$ & $1.61\times10^{23}$ & $4.80\times10^{26}$  \\
     2.0  & $3.66\times10^{26}$ & $1.54\times10^{24}$ & $1.14\times10^{23}$ & $3.68\times10^{26}$  \\
     2.2  & $2.88\times10^{26}$ & $1.02\times10^{24}$ & $9.16\times10^{22}$ & $2.89\times10^{26}$ \\
     2.4  & $2.30\times10^{26}$ & $6.69\times10^{23}$ & $7.68\times10^{22}$ & $2.30\times10^{26}$  \\
     2.6  & $2.19\times10^{26}$ & $7.70\times10^{23}$ & $7.98\times10^{22}$ & $2.19\times10^{26}$  \\
     2.8  & $1.67\times10^{26}$ & $1.13\times10^{24}$ & $5.29\times10^{22}$ & $1.68\times10^{26}$  \\
     3.0  & $1.51\times10^{26}$ & $8.94\times10^{23}$ & $4.10\times10^{22}$ & $1.52\times10^{26}$  \\
     3.2  & $1.04\times10^{26}$ & $1.19\times10^{24}$ & $4.13\times10^{22}$ & $1.05\times10^{26}$  \\
     3.4  & $1.10\times10^{26}$ & $1.21\times10^{24}$ & $3.69\times10^{22}$ & $1.11\times10^{26}$  \\
     3.6  & $6.57\times10^{25}$ & $6.58\times10^{23}$ & $1.57\times10^{22}$ & $6.64\times10^{25}$ \\
     3.8  & $9.13\times10^{25}$ & $6.14\times10^{23}$ & $2.81\times10^{22}$ & $9.19\times10^{25}$  \\
     4.0  & $8.69\times10^{25}$ & $7.47\times10^{23}$ & $2.64\times10^{22}$ & $8.77\times10^{25}$  \\
		\hline
		\hline
	\end{tabular}
	
\end{table}

\begin{figure*}
	\centering 
	\includegraphics[width=\textwidth]{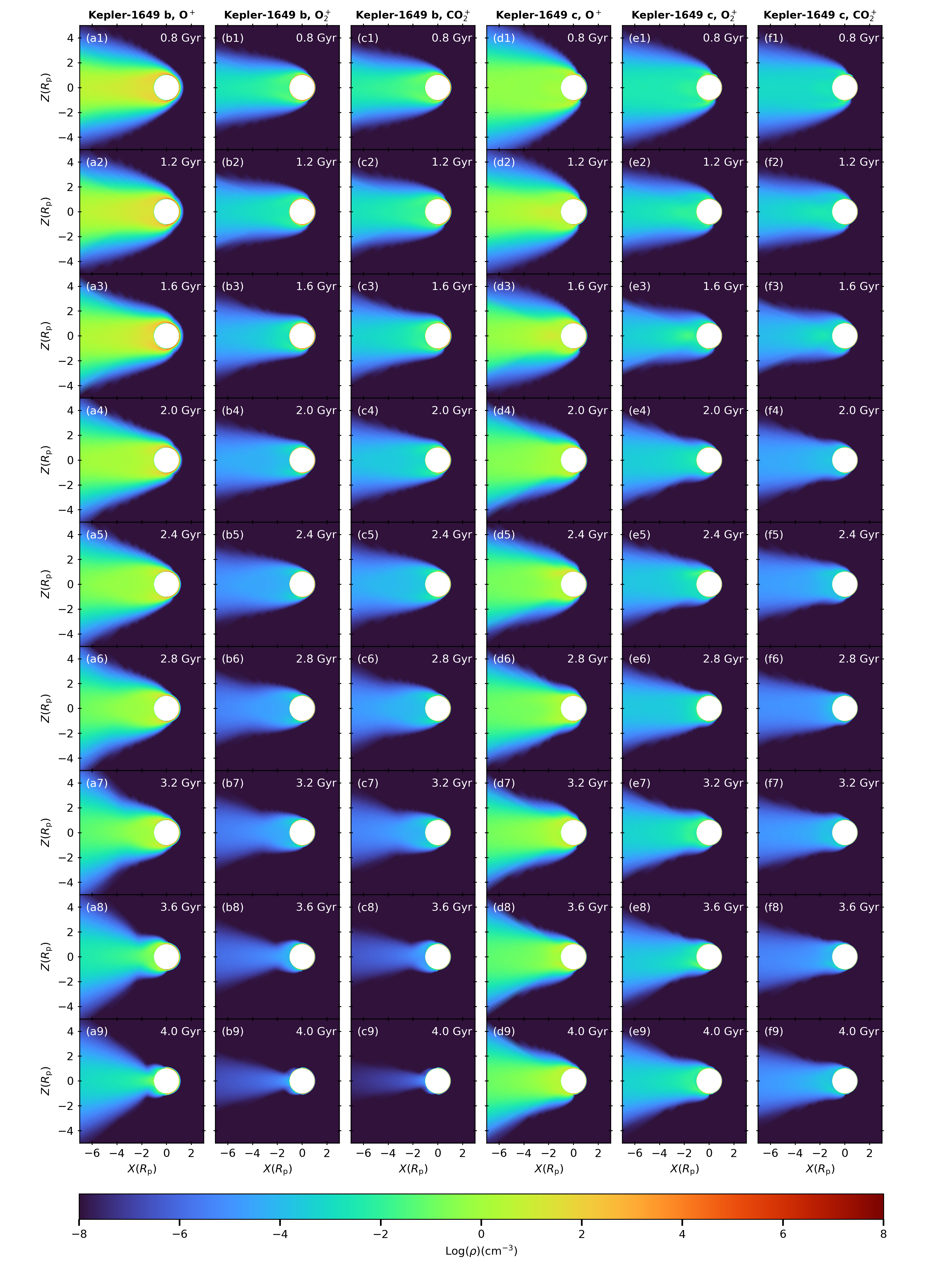}
	\caption{Logarithmic ion number density (cm$^{-3}$) distributions in the X-Z plane for Kepler-1649 b and c at different stellar ages (0.8-- 4.0 Gyrs; top to bottom). Columns show \ce{O+}, \ce{O2+}, and \ce{CO2+} (left to right). Coordinates are normalized to planetary radius $R_{\rm P}$.
        }
	\label{Fig:2D_distribution_Kb_c}
\end{figure*}

Figures~\ref{Fig:2D_distribution_Kb_c} illustrate the temporal evolution of ion escape patterns for Kepler-1649 b and c across 0.8--4.0 Gyr, highlighting two key trends: (1) significantly higher ion escape rates in the early evolutionary stages due to intense stellar winds and XUV radiation, and (2) the dominance of \ce{O+} in the escaping ion flux across all epochs. Both planets exhibit systematically higher ion densities and more extended ionospheres at younger stellar ages, particularly for \ce{O+}, which dominates the escaping flux. The dominance of \ce{O+} escape is visually evident in the extended ion tails in Figures~\ref{Fig:2D_distribution_Kb_c}. This early high escape is supported by the dense ionospheric structures (see Figures \ref{Fig:Kepler1649b_c_Oion} in the Appendix), where elevated dayside \ce{O+} densities enhance ion loss, driven by the strong XUV flux and stellar wind pressure in the early epochs. Comparative analysis shows that Kepler-1649 b exhibits higher dayside ionospheric densities than Kepler-1649 c, driven by stronger stellar radiation due to its closer orbital distance. These findings highlight the critical influence of orbital proximity (0.051 AU for Kepler-1649 b vs. 0.088 AU for Kepler-1649 c) and stellar evolution in regulating atmospheric loss. Overall, the results demonstrate that the atmospheric erosion histories of M-dwarf planets are shaped by both orbital distance and the evolving stellar wind and radiation environment.

\section{Discussion and Conclusion} \label{sec:discussion}

This work investigates the long-term atmospheric retention of Kepler-1649 b and c throughout the evolutionary history of their M-dwarf host, tackling a key challenge in assessing exoplanet habitability. We employed a well-validated multi-species magnetohydrodynamic (MHD) model to simulate time-dependent ion escape driven by stellar winds and extreme ultraviolet (EUV) radiation across 0.8--4.0 Gyr. Our focus on these Earth-sized planets offers a unique opportunity to probe atmospheric retention under M-dwarf conditions, where stellar activity strongly influences planetary evolution. Expanding on previous research, this study provides an evolutionary context for atmospheric loss, yielding fundamental insights into the persistence of terrestrial exoplanet atmospheres orbiting active M-dwarfs.

The decay model captures the dynamical evolution of atmospheric escape in active M-dwarf systems. In the Kepler-1649 system, the observed decay of total ion escape rates (Fig.~\ref{fig1}) arises from the decline of both stellar wind dynamic pressure ($P_{\rm dyn} \propto \tau^{-2.1}$) and XUV flux ($F_{\rm XUV} \propto \tau^{-0.7}{\sim} \tau^{-1.5}$), consistent with rotational braking models calibrated for M-dwarfs like Kepler-1649 \citep{2017A&A...603A..58R}. Correspondingly, escape rates decrease from $4.49 \times 10^{27}~\rm s^{-1}$ and $1.20 \times 10^{27}~\rm s^{-1}$ at 0.8 Gyr to $2.22 \times 10^{24}~\rm s^{-1}$ and $8.77 \times 10^{25}~\rm s^{-1}$for Kepler-1649 b and c, respectively, at 4.0 Gyr. The decline in ion escape rates from 0.8 to 4.0 Gyr highlights the critical influence of stellar evolution on the atmospheric retention of Kepler-1649 b and c. This trend aligns with the expected weakening of stellar activity in M-dwarfs, where early epochs are marked by intense high-energy radiation and particle fluxes that drive significant atmospheric erosion, while later stages allow for increased atmospheric stability \citep{2010Icar..210..539Z, 2020AJ....160..237F}. The decline in ion escape rates is physically driven by the interplay of stellar wind dynamic pressure, which decays as $\propto \tau^{-2.1}$ in our model due to rotational braking and reduced mass-loss rates in aging M-dwarfs \citep{Garraffo2016}, and XUV flux evolution, which heats the upper atmosphere, drives thermal expansion, and enhances photoionization source terms for escaping ions \citep{lammer2008atmospheric}. For Venus-like orbits like Kepler-1649~b, the higher stellar wind dynamic pressure (up to $2.7\times$ that of planet~c) directly compresses the ionosphere, accelerating ion pickup and forming extended escape tails, consistent with MHD studies of unmagnetized exoplanets \citep{ma2013global,dong2018atmospheric}. Early stellar activity amplifies this, with $F_{\rm XUV}$ several orders of magnitude above solar levels, boosting $q_{\ce{O}}$ and $q_{\ce{CO2}}$ and leading to \ce{O+} dominance (98.3\%--99.9\% of loss) via efficient photochemical chains: \ce{CO2+ + O -> O2+ + CO} (followed by dissociative recombination). Atmospheric composition further modulates this: in \ce{CO2}-\ce{O} dominated setups, neutral densities and reaction rates e.g., $\alpha_{\rm rec} = 3.1e-7\, (300/T_e)^{0.5}$ cm$^3$\,s$^{-1}$ favor \ce{O+} escape over lighter species, differing from H-dominated atmospheres where hydrodynamic blowoff might prevail. Orbital distance geometrically scales both parameters: $P_{\rm dyn} \propto 1/r^2$  and $F_{\rm XUV} \propto 1/r^2$ explaining b's 3.79$\times$ higher \ce{O+} escape rates at the early epoch. Absent a global magnetic field (as assumed for these Venus-analogs), stellar winds interact directly with the ionosphere, maximizing erosion. These mechanisms, validated against Venus-like analogs \citep{dong2018atmospheric}, underpin the robustness of nonparametric regression trends within Kepler-1649's parameter space. Synthetic JWST transmission spectra shown in Fig. \ref{Fig:JWSTspectra}, informed by our escape simulations, suggest observable evolutionary signatures (Section \ref{sec:JWSTspectra} in the Appendix). The synthetic spectra demonstrate JWST could test the escape trends statistically across similar systems, constraining upper mass-loss limits through evolutionary spectral shifts rather than direct detection \citep{2016MNRAS.458.2657B,2019AJ....158...27L}.

The differences in atmospheric ion escape rates between Kepler-1649 b and c arise from their orbital separations and the evolving stellar wind and XUV radiation environment. At 0.8 Gyr, Kepler-1649 b (0.051 AU) exhibits a total ion escape rate of $4.49 \times 10^{27}~\rm s^{-1}$, 3.74 times higher than Kepler-1649 c's $1.2 \times 10^{27}~\rm s^{-1}$ (0.088 AU), a disparity rooted in the inverse-square scaling of stellar wind dynamic pressure and XUV flux \citep{2017A&A...603A..58R}. However, in later epochs (beyond $\sim$2.4 Gyr), Kepler-1649 b's escape rates fall below those of Kepler-1649 c, with the reversal most pronounced at 4.0 Gyr; this unexpected trend is attributable to a shift in the star-planet interaction regime toward sub-magnetosonic conditions, as evidenced by declining fast magnetosonic Mach numbers ($M_{\rm f} < 1.1$ for the last four evolutionary stages, reaching $M_{\rm f} \approx 0.964$ at 4.0 Gyr as listed in Table~\ref{table1}). In this sub-magnetosonic regime, analogous to Ganymede's magnetosphere embedded in Jupiter's plasma flow \citep{2018JGRA..123.2815W}, an upstream bow shock does not form; instead, Alfvén-wing-type structures develop, which limit the efficiency of stellar wind-driven atmospheric stripping and reduce atmospheric ion escape rates despite Kepler-1649 b's proximity to the star \citep{2018PNAS..115..260D}. Based on the calculation, the cumulative \ce{O+} loss for Kepler-1649 b reaches $7.65 \times 10^{43}$ ions (equivalent to 0.32 bar) over the 0.8–4.0 Gyr period. This suggests that both Kepler-1649 b and Kepler-1649 c—with the latter experiencing approximately half the integrated loss—could retain a 1\,bar \ce{CO2}-dominated atmosphere for several billion years. These findings imply that Kepler-1649 b and c may be capable of sustaining atmospheres conducive to long-term habitability. Retention depends on initial \ce{CO2}-\ce{O} dominance; \ce{N2}-rich cases might reduce \ce{O+} escape via altered photochemistry, while \ce{H2O}-rich could amplify loss via \ce{H+} channels. This study builds on prior work by quantifying time-dependent escape throughout the evolutionary history of a star. Our nonparametric regression trends for Kepler-1649 b and c provide system-specific insights into atmospheric loss in the Kepler-1649 b and c and may offer a reference for similar M-dwarf systems. While our nonparametrically regressed trends are derived solely from Kepler-1649 and should not be extrapolated universally, they may hold for other M-dwarf systems with comparable stellar activity profiles and Venus-like atmospheric compositions, pending further studies.

We note that our model assumes fixed circular orbits without planet-planet interactions. However, even if both orbits are initially 
circular, they are going to become eccentric \citep[e.g.,][]{2003MNRAS.345..340G,2009MNRAS.392.1253G}, thus altering star-planet distances and possibly atmospheric erosion rates. Although the planetary eccentricities of the specific system are not known, some constraints, however, can arise from the fact that the system must be dynamically stable.  Assuming initially circular orbits for both Kepler-1649 b and c, the stability results of hierarchical triple systems given in \cite{2013NewA...23...41G} indicate that a system such as Kepler-1649 would be fairly stable. Nonetheless, if any of the planets is on a mildly eccentric orbit, both secular and resonant oscillations (our system is close to a 9:4 mean motion resonance) may become more significant for the dynamical evolution the system. Assuming a circular orbit for Kepler-1649b, \cite{2021AJ....161...31K} found that the system became unstable when the eccenticity of Kepler-1649c had an initial eccentricity beyond 0.325 in their numerical experiments. Future work should address how orbital distance variability impacts long-term atmospheric retention in this system. 

Future investigations should also incorporate dynamic atmospheric models and additional physical processes to enhance the model’s predictive power. Adopting time-evolving photochemical models \citep{2017ApJS..228...20T,cangi_2024_11307881} would capture neutral atmosphere evolution, addressing the static assumption’s shortcomings. Including planetary magnetic field evolution, as in \citet{10.1093/mnras/stab2947}, could reveal shielding effects on ion escape. Three-dimensional stellar wind models \citep{2022ApJ...941L...8G, Cohen_2023} would better resolve spatial variability, while integrating secondary ionization \citep{2023A&A...680A..33G}, kinetic processes \citep{2005JGRA..110.3221S} and outgassing \citep{2007Icar..186..462S,2016ApJ...828...80K,2022ARA&A..60..159W,2024ApJ...960...44T} would provide a holistic view of atmospheric budgets. Constraining Kepler-1649’s age with multi-method stellar chronology (e.g., gyrochronology, isochrones) would further anchor the evolutionary timeline, refining habitability predictions for these exoplanets.

In summary, this Letter highlights several key findings. The atmospheric ion escape rates for Kepler-1649 b and c exhibit decline trends with stellar age. Both planets appear capable of retaining their atmospheres over 4.0 Gyr, carrying important implications for their potential habitability. The results emphasize the crucial role of orbital distance and stellar evolution in regulating atmospheric retention and evolution in the Kepler-1649 system, offering a benchmark for interpreting upcoming JWST observations of similar M-dwarf exoplanets.

\section*{Acknowledgements}
The authors are grateful to the anonymous referee for the valuable comments and suggestions which helped to improve the paper. This work was supported by the State Key Laboratory of High Temperature Gas Dynamics (Grant No. 2023KF02), the National Natural Science Foundation of China (Grant No. 12273043), the National Laboratory on Adaptive Optics, China (Grant No. FNLAO-24-MS-O09), the China's Space Origins Exploration Program (Grant No. GJ11030206), the Youth Innovation Promotion Association CAS (Grant Nos. 2018178, 2021144), the China Scholarship Council (Grant No. 202404910139). H.-L. Y acknowledges the National Natural Science Foundation of China with grant No. 12373036, 12022304, and the support from the Youth Innovation Promotion Association of the Chinese Academy of Sciences. This research was also partially supported by the Munich Institute for Astro-, Particle and BioPhysics (MIAPbP), which is funded by the Deutsche Forschungsgemeinschaft (DFG, German Research Foundation) under Germany´s Excellence Strategy – EXC-2094 – 390783311. M.T.M. acknowledges funding from the Bell Burnell Graduate Scholarship Fund, administered and managed by the Institute of Physics, and the Croucher Fellowship funded by the Croucher Foundation. NM acknowledges support from a UKRI Future Leaders Fellowship [grant number MR/T040866/1]. C.D. acknowledges support from the Alfred P. Sloan Research Fellowship. We thank for the technical support of the National Large Scientific and Technological Infrastructure ``Earth System Numerical Simulation Facility  (EarthLab)'' (\url{https://cstr.cn/31134.02.EL}). The numerical computations in this paper were conducted on the Hefei Advanced Computing Center and EarthLab. The Space Weather Modeling Framework that comprises the BATS-R-US code used in this study is publicly available at \url{https://github.com/SWMFsoftware/BATSRUS}.  

%\facilities{ADS, NASA Exoplanet Archive.}
\software{\texttt{numpy} \citep{harris2020array},
	\texttt{scipy} \citep{2020SciPy-NMeth},
	\texttt{matplotlib} \citep{2007CSE.....9...90H}.}

\appendix

\section{Parker's stellar wind model} \label{sec:Parker_wind}

The Parker's wind equation is read as \citep{priest2012solar,griessmeier2006aspects}
\begin{equation}
	\left(\frac{v(d)}{v_{\rm crit}}\right)^2 - 2{\rm ln}\left(\frac{v(d)}{v_{\rm crit}}\right)^2 = 4{\rm ln}\frac{d}{r_{\rm crit}} + 4\frac{r_{\rm crit}}{d} - 3.
	\label{eq1}
\end{equation}
By solving this equation, the stellar wind velocity $v(d)$ at distance $d$ can be obtained.
The critical velocity $v_{\rm crit}$ is defined as
\begin{equation}
	v_{\rm crit} = \sqrt{\frac{k_{\rm B}T}{m}},    
	\label{eq2}
\end{equation}
and the critical radius $r_{\rm crit}$ is
\begin{equation}
	r_{\rm crit} = \frac{mGM_*}{4k_{\rm B}T}.
	\label{eq3}
\end{equation}
We can also calculate the density $n(r)$ of the stellar wind by
\begin{equation}    
	n(r)=\frac{\dot{M}_*}{4\pi d^2v(d)m}.
	\label{eq4}
\end{equation}
Here, $k_{\rm B}$ is the Boltzmann constant, $T$ is the temperature of the stellar wind, $m$ is the mass of the stellar wind protons, $G$ is the gravitational constant, and $\dot{M}_*=4\pi d^2v(d)m$ is the stellar mass loss rate.
To validate these formulas, we perform calculations for the Sun's stellar wind velocity and number density at 1 AU. The results yield approximately 422 km/s and 6.62 $\rm cm^{-3}$, respectively, which are consistent with observations obtained from spacecrafts such as SOHO and vela-3. It is noteworthy that the method in \cite{griessmeier2006aspects} limits a stellar age greater than 0.7 Gyr. For younger stars, the stellar wind dynamic pressure is actually much stronger than that obtained by this method, which is why we chose 0.8 Gyr as the lower limit on the time point.

\cite{griessmeier2006aspects} provided a method for calculating solar wind parameters of M-type stars based on the formula mentioned above. Using a set of M-type stars with a standard orbital radius of 1 AU and a stellar age of 4.6 Gyr, we iteratively adjust the coronal temperature until the velocity derived from Eq. (\ref{eq1}) matches the standard value. Subsequently, we substitute the obtained coronal temperature into Eqs. (\ref{eq2}) and (\ref{eq3}) to calculate the stellar wind speed for the target radius and stellar age. Finally, we use the obtained velocity in Eq. (\ref{eq4}) to compute the density. The calculated results are listed in Table \ref{table1}. Within 1 AU, as the distance increases, the stellar wind density decreases while the velocity increases, which is consistent with the results from \cite{griessmeier2006aspects}.

The IMF is calculated with the equations \citep{parker1958dynamics}

\begin{align}
	B_r(r,\theta,\phi)&=B(\theta,\phi_0)\left(\frac{b}{r}\right)^2, \\
	B_{\theta}(r,\theta,\phi)&=0, \\
	B_{\phi}(r,\theta,\phi)&=B(\theta,\phi_0)\left(\frac{\omega}{v_m}\right)(r-b)\left(\frac{b}{r}\right)^2 {\rm sin}(\theta).
\end{align}

These equations use a spherical coordinate system centered on the star to describe the decay of the IMF with the $1/r^2$, where $r$ is the distance between the target position and the star, $b$ is a distance beyond which the solar gravitation and outward acceleration by high coronal temperature are neglected, $B(\theta,\phi_0)$ is the magnetic field at $r=b$, $\omega$ is the angular velocity of Kepler-1649, which can calculate by empirical relations from \cite{2023MNRAS.525.5168M}. If only the dipole magnetic field of the star is considered, $B(\theta,\phi_0)$ can be considered as the vertical component of the surface magnetic field at the planet. By substituting the calculated stellar wind velocities obtained from the previous section into the equations, the stellar wind magnetic field parameters for different stellar ages are obtained.

\section{Stellar spectra} \label{sec:Stellar_spectra}

\begin{figure*}[h!]
	\centering 
	\resizebox{\hsize}{!}
	{\includegraphics[width=0.6\textwidth]{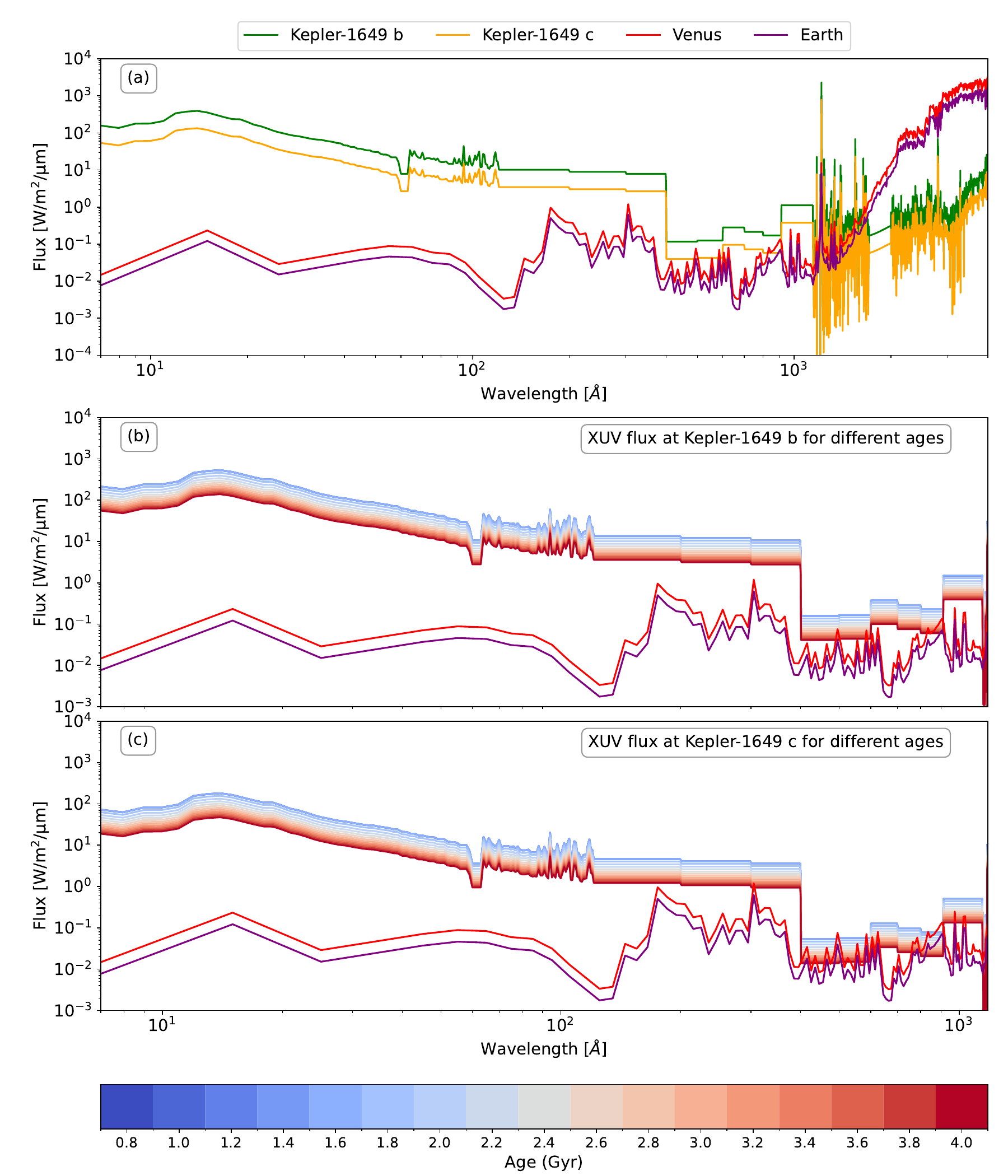}}
	\caption{Stellar spectra at Kepler-1649 b and c, and also Solar spectra at Venus and Earth for comparison. Panel (a): flux (W~m$^{-2}$~$\mu$m) for Kepler-1649 at the orbital distances of planets b (0.0514\,AU, green line) and c (0.0882\,AU, orange line), compared to the solar spectrum at Venus (red line) and Earth (purple line). Kepler-1649's spectrum is scaled from Proxima Centauri (M5.5V), matching its spectral type. Panels (b) and (c): the temporal evolution of the XUV flux (0.7 nm--118 nm, W m$^{-2}$~$\mu$m) for Kepler-1649 at the orbital distances of planets b and c, respectively, spanning stellar ages of 0.8 to 4.0 Gyr.}
	\label{Fig:Kepler1649bc_spec}
\end{figure*}

Figure~\ref{Fig:Kepler1649bc_spec} presents the stellar spectra at the orbital distances of Kepler-1649 b and c, compared with solar spectra at Venus and Earth, along with the temporal evolution of XUV flux. This evolutionary pattern drives the corresponding decrease in photoionization rates that directly influences atmospheric escape processes simulated in our MHD models. Figure 4 shows $F_{\rm XUV}$ versus stellar age ($\tau$) for Kepler-1649. The profile illustrates the early saturation phase and subsequent power-law decay, normalized to Kepler-1649 b and c orbital distances. This average flux excludes flares which are not modeled here, as our focus is on long-term ion escape trends. 

\newpage

\begin{figure*}[h!]
	\centering 
	\resizebox{\hsize}{!}
	{\includegraphics[width=0.4\textwidth]{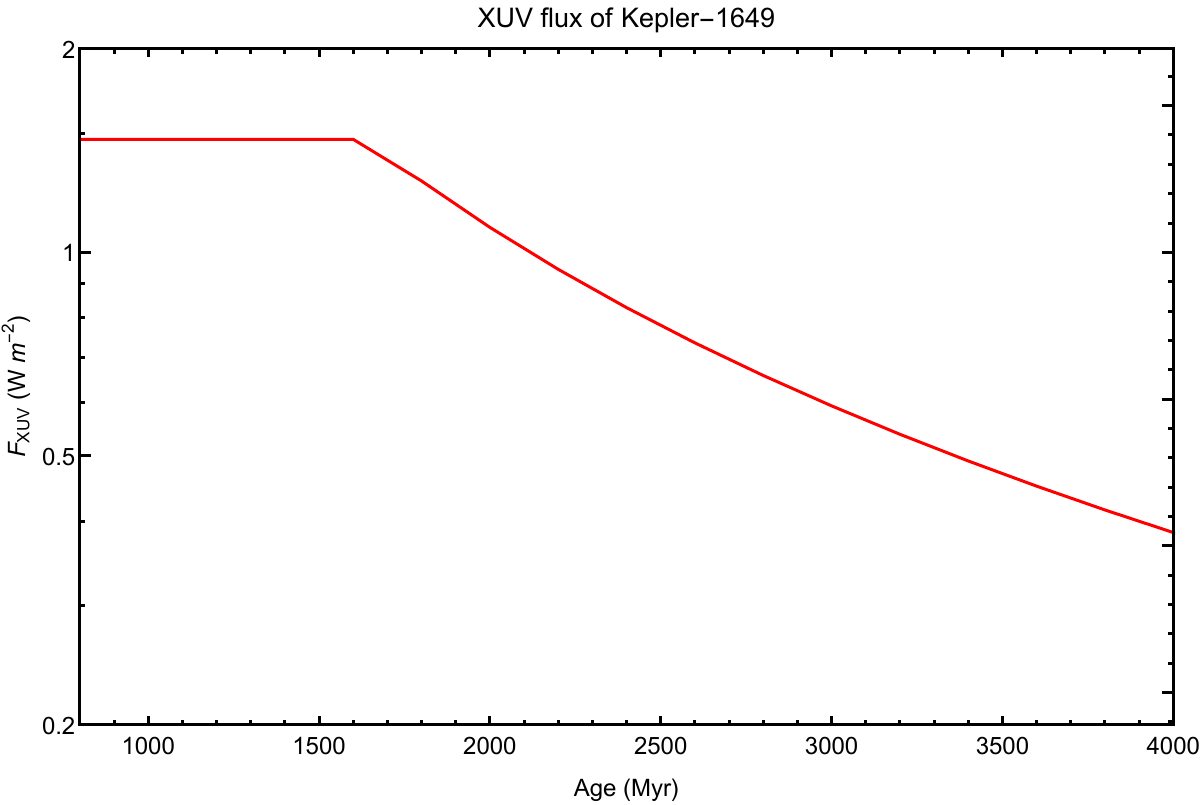}}
	\caption{The $F_{\rm XUV}$ profile with time for Kepler-1649, spanning stellar ages of 0.8 to 4.0 Gyr.}
	\label{Fig:Kepler1649_FXUV}
\end{figure*}

\section{Chemical Reactions and related rates} \label{sec:chemical_reactions}

We calculated the photoionization rates using the equation \citep{1985JGR....90.6675T}
\begin{gather}
q_i = \int_{\lambda} \sigma_i(\lambda) F_{\infty}(\lambda)d\lambda,
\end{gather}
where $\sigma_i$ is the ionization cross-section and $F_{\infty}$ is the radiative flux at the top of the atmosphere. By integrating the product of these quantities over wavelength, we determined the photoionization rates as input parameters.

\begin{table*}[h]
%    \tablenum{2}
    \centering
    \caption{Chemical Reactions and related rates.}\label{table3}
    \begin{tabular}{lll}
    \hline
    \hline
    \multicolumn{2}{c} {Chemical Reaction} & Rate Coefficient\footnote{The reaction rates are based on \cite{schunk2009}, with electron impact ionization omitted in the calculations. The $\mathrm{H}^+$ density is sourced from the stellar wind, and neutral hydrogen is disregarded.} \\
    \hline
    \multicolumn{3}{c} {Primary Photolysis\footnote{The photoionization rates are derived and scaled to correspond to Kepler-1649 b and c, employing the EUV flux calculated through Eq. (\ref{eq_XUV}).} in s$^{-1}$} \\
    \hline
    & \ce{CO2} + $h\nu$ $\rightarrow$  \ce{CO2+} + \ce{e-} &  see Table \ref{table1}   \\ %(solarmin)$
    & \ce{O} + $h\nu$ $\rightarrow$  \ce{O+} + \ce{e-}     &  see Table \ref{table1}  \\
    \hline
    \multicolumn{3}{c} {Ion-neutral and electron recombination chemistry in cm$^3$\,s$^{-1}$} \\
    \hline    
    &  \ce{CO2+} + \ce{O} $\rightarrow$  \ce{O2+} + \ce{CO}  & $1.64 \times 10^{-10}$   \\
    &  \ce{CO2+} + \ce{O} $\rightarrow$  \ce{O+} + \ce{CO2}  & $9.60 \times 10^{-11}$  \\
    &  \ce{O+} + \ce{CO2} $\rightarrow$  \ce{O2+} + \ce{CO}  & $1.1 \times 10^{-9}$ (800/T$_i$)$^{0.39}$  \\
    &  \ce{H+} + \ce{O} $\rightarrow$  \ce{O+} + \ce{H}\footnote{The rate coefficient is adopted from \cite{fox01}.}  & $5.08 \times 10^{-10}$  \\
    &  \ce{O2+} + \ce{e-} $\rightarrow$  \ce{O} + \ce{O}  & $7.38 \times 10^{-8}$ (1200/T$_e$)$^{0.56}$ \\
    &  \ce{CO2+} + \ce{e-} $\rightarrow$  \ce{CO} + \ce{O}  & $3.10 \times 10^{-7}$ (300/T$_e$)$^{0.5}$ \\
    \hline
    \hline
    \end{tabular}
% \textit{NOTE--- a)} Electron impact ionization is neglected in the calculation, H$^+$ density is from the stellar wind, the neutral hydrogen is neglected.

\end{table*}

\section{\ce{O+} and \ce{O2+} distribution in the dayside ionosphere at 0.8 Gyr and 4.0 Gyr} \label{sec:distribution}

\begin{figure*}[h!]
    \centering 
    \resizebox{\hsize}{!}
    {\includegraphics[width=0.4\textwidth]{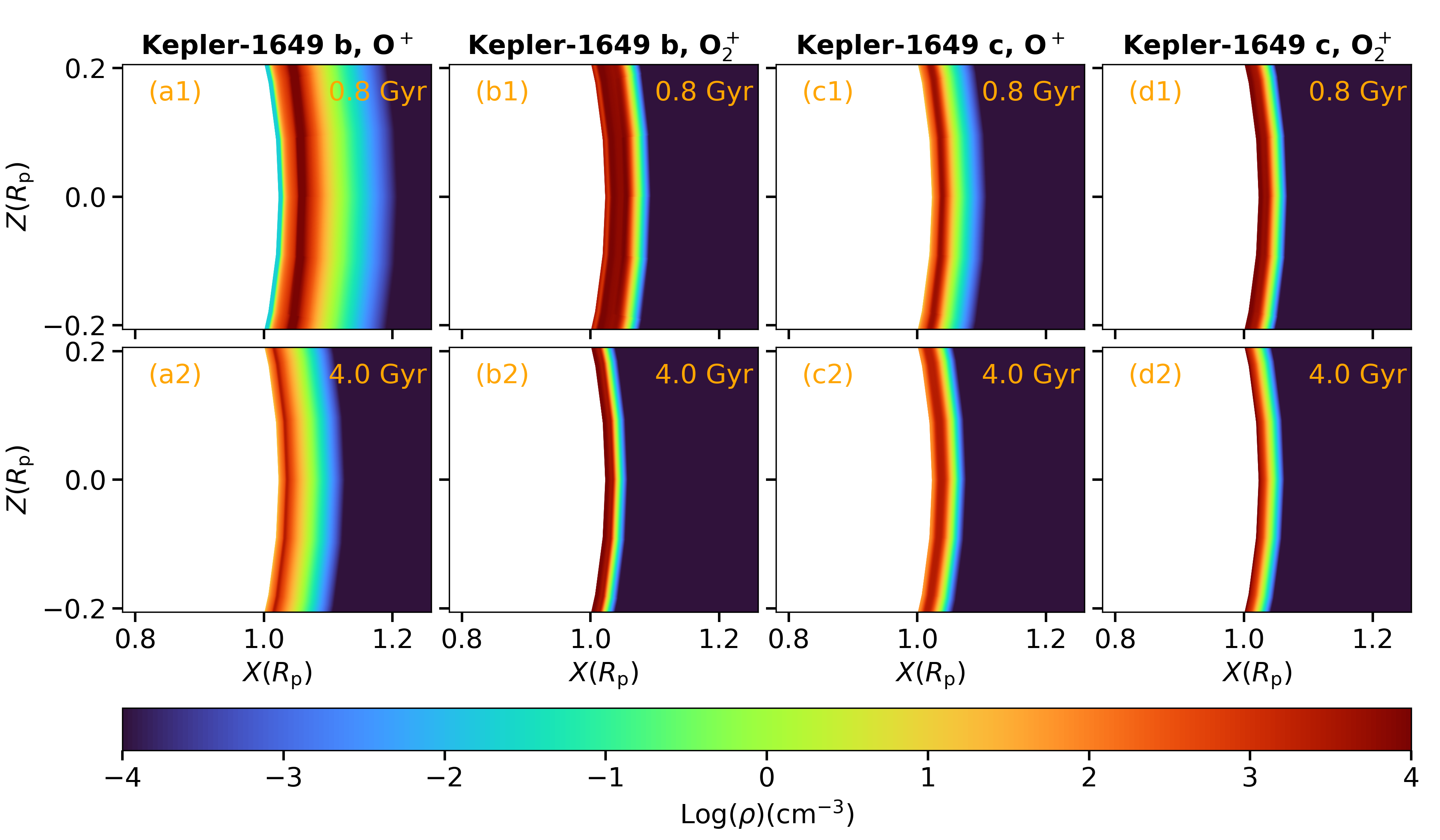}}
    \caption{Logarithmic ion number density (cm$^{-3}$) distributions of \ce{O+} (left column) and \ce{O2+} (right column) in the X-Z plane for Kepler-1649 b and c's dayside at 0.8 Gyr (top row) and 4.0 Gyr (bottom row). Coordinates are normalized to the planetary radius $R_{\rm P}$.}
    \label{Fig:Kepler1649b_c_Oion}
\end{figure*}

The logarithmic distributions of \ce{O+} and \ce{O2+} in the dayside ionosphere of Kepler-1649 b and Kepler-1649 c at 0.8 Gyr and 4.0 Gyr are shown in Figure \ref{Fig:Kepler1649b_c_Oion}. The plots illustrate the spatial variation of \ce{O+} and \ce{O2+}, highlighting the evolution of ionospheric ion distribution over time. 

\newpage

\newpage
\section{Synthetic JWST spectra} \label{sec:JWSTspectra}

While our conclusions are system-specific to Kepler-1649 and should not be universally extrapolated, future discoveries will enable statistical tests of atmospheric escape evolution. With growing samples of exo-Venuses from TESS and PLATO, we can construct subsets of systems similar to Kepler-1649 in stellar activity, orbital distances, and atmospheric compositions (e.g., CO2-O dominated). By comparing transmission spectra across ages (0.8-4.0 Gyr), trends in escape rates could be observationally validated, inferring mass-loss upper limits from changes in scale height or composition (e.g., O+ depletion implying reduced escape). 

\begin{figure*}[h!]
    \centering 
    \resizebox{\hsize}{!}
    {\includegraphics[width=0.4\textwidth]{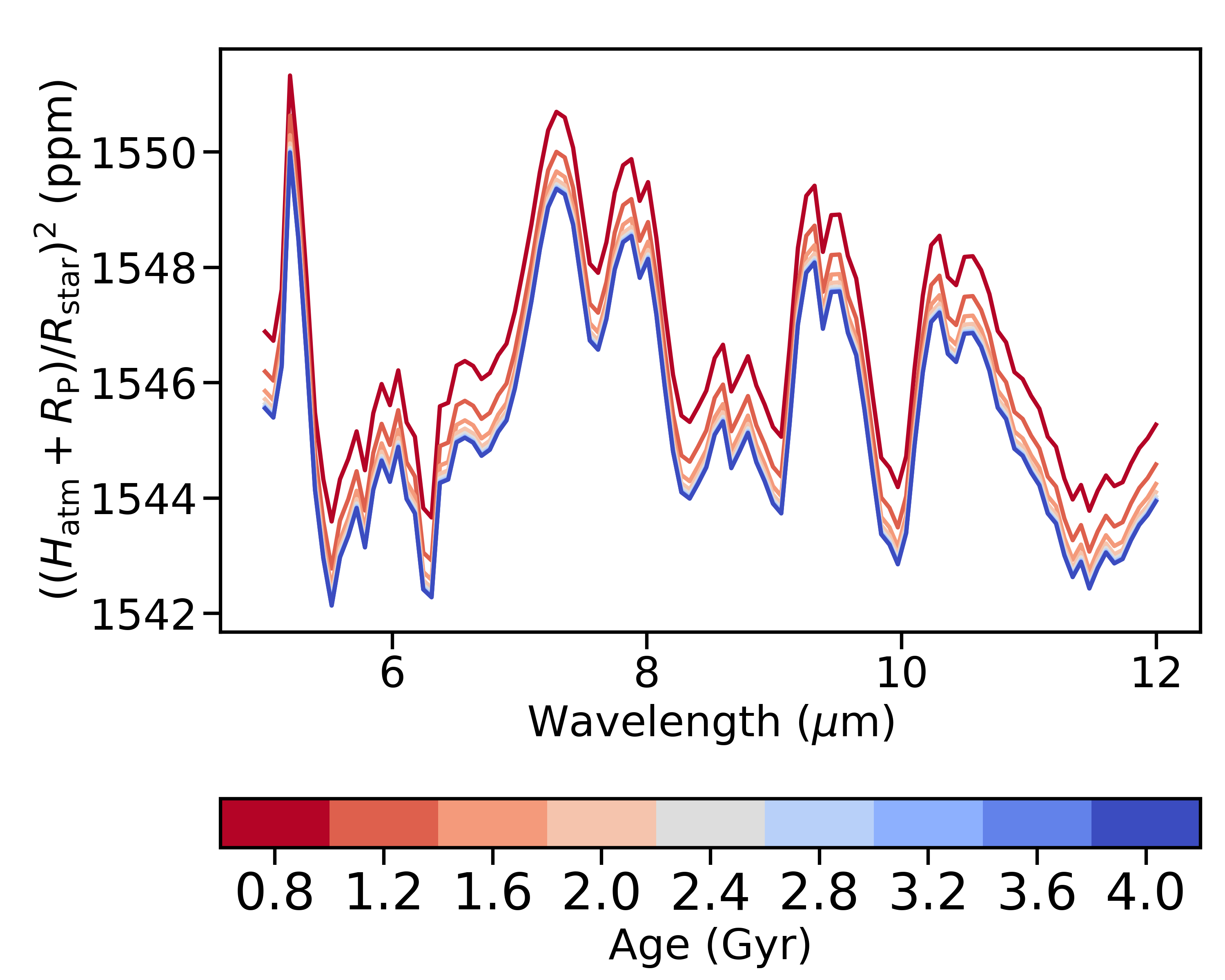}}
    \caption{The prediction of evolution of transmission spectra for Kepler-1649 b at nine key evolutionary ages from 0.8-4.0 Gyr.  $R_{\rm p}$ is the solid radius of planet, $H_{\rm atm}$ is the opaque component of the atmosphere, and $R_{\rm star}$ is the radius of star.}
    \label{Fig:JWSTspectra}
\end{figure*}

To illustrate this, we generated synthetic transmission spectra for Kepler-1649 b at each evolutionary stage using the Planetary Spectrum Generator (PSG; \citep{VILLANUEVA201886}), assuming a Venus-like atmosphere modulated by our simulated ion escape rates (e.g., reduced \ce{O} and \ce{CO2} densities at later ages). The spectra (Fig. \ref{Fig:JWSTspectra}) show evolving transit depths, with stronger \ce{CO2} features (2.7-4.3 $\mu$m) at early epochs (higher escape) diminishing over time as the atmosphere stabilizes.

\newpage

\newpage

\bibliography{sample631}{}
\bibliographystyle{aasjournal}



\end{document}